\documentclass[%
 reprint,
 amsmath,amssymb,
 aps,
]{revtex4-1}

\usepackage{graphicx}
\usepackage{dcolumn}
\usepackage{bm}

\begin{document}

\preprint{APS/123-QED}

\title{Characterization of charge-exchange collisions between ultracold $\rm{^6Li}$ atoms and $\rm{^{40}Ca^+}$ ions}

\author{R. Saito$^{1}$}
\author{S. Haze$^{1}$}%
\author{M. Sasakawa$^{1}$}
\author{R. Nakai$^{1}$}
\author{M. Raoult$^{2}$}
\author{H. Da Silva, Jr.$^{2}$}
\author{O. Dulieu$^{2}$}
\author{T. Mukaiyama$^{1,3}$}
\email{Email address : muka@ils.uec.ac.jp}
\affiliation{
$^{1}$\mbox{Institute for Laser Science,University of Electro-Communications, Chofugaoka, Chofu, Tokyo 182-8585, Japan}\\
$^{2}$\mbox{Laboratoire Aim\'e Cotton, CNRS, Universit\'e Paris-Sud, ENS Cachan, Universit\'e Paris-Saclay, 91405 Orsay Cedex, France}\\
$^{3}$\mbox{PRESTO, Japan Science and Technology Agency, Honcho, Kawaguchi, Saitama 332-0012 Japan}\\
}



\date{\today}

\begin{abstract}
We investigate the energy dependence and the internal-state dependence of the charge-exchange collision cross sections in a mixture of $^6$Li atoms and $^{40}$Ca$^+$ ions.
Deliberately excited ion micromotion is used to control the collision energy of atoms and ions.
The energy dependence of the charge-exchange collision cross section obeys the Langevin model in the temperature range 
of the current experiment, and the measured magnitude of the cross section is correlated to the internal state of the $^{40}$Ca$^+$ ions.
Revealing the relationship between the charge-exchange collision cross sections and the interaction potentials is 
an important step toward the realization of the full quantum control of the chemical reactions at an ultralow temperature regime.
\end{abstract}

\pacs{Valid PACS appear here}
\maketitle



\section{\label{sec:level1}Introduction}

A system of laser-cooled ions immersed in an ultracold atomic gas introduces a new experimental degree of freedom, a charge, into cold atom experiments \cite{Winthrop, Grier, Zipkes2, Zipkes, Schmid, Rellergert, Hall, Hall2, Sullivan, ratschbacher2012controlling, Ravi, Haze, Lee, Ratschbacher2, Haze2}. 
A single ultracold ion would be useful as a local probe of an ultracold atomic cloud with high spatial resolution, since ions can be controlled using an electric field without exerting much influence on the atoms \cite{Sherkunov, Kollath}.
A system of several crystalized ions mixed with neutral atoms would nicely model atomic cores comprising a lattice mixed with an electron gas in a solid-state material \cite{Bissbort}. The charge of the ions would activate the atoms and ions to chemical reactions, and therefore the atom-ion hybrid system would make an ideal platform to study chemical reactions at a single atom level.

When atoms and ions are trapped together, the ions attract the atoms by their static electric field and sometimes lead to inelastic collisions.
So far, charge-exchange collisions either with or without emission of radiation \cite{Schmid, Zipkes, Rellergert, Sullivan, Hall, Haze2} and molecular association due to radiative processes \cite{Hall, Humberto, Hall2, Hall3} or three-body collisions \cite{Harter2, Krukow, Krukow2} have been studied in ultracold atom-ion hybrid systems. 
A systematic study of such inelastic collisions, which are elementary chemical reaction processes, is essential to realize the full quantum control of the chemical reactions in an ultralow temperature regime. The knowledge of the molecular interaction potentials and the potential couplings is important for understanding such chemical reactions.
The experimental characterization of the ultracold inelastic collisions under the precise control of the collision energies and the internal states of the reacting atoms and ions allows for testing the interaction potentials which are obtained from the elaborated quantum chemistry calculations.

In this article, we systematically study the energy dependence and the internal-state dependence of the charge-exchange cross sections in an ultracold $^6$Li--$^{40}$Ca$^+$ mixture. Our choice of the atom-ion combination is unique and promising for the observation of quantum collisions since the $^6$Li--$^{40}$Ca$^+$ combination has a high s-wave threshold energy of 10~$\mu$K and a quite low limit of the micromotion-induced heating rate of a few microkelvin per second~\cite{Marko}. To control the collision energies precisely, we deliberately excite an excess ion micromotion in a linear Paul trap. We determine the energy of the ions from the micromotion-modulated fluorescence spectra \cite{Devoe, Blumel, Berkeland}.
The energy dependence of the charge-exchange cross section $\sigma_{\rm CE} (E)$ is scaled with the energy-dependent Langevin cross section $\sigma_{\rm L} (E)$ and the cross section can be expressed as $\sigma_{\rm CE} (E) = A \sigma_{\rm L} (E)$. The factor $A$ describes the probability of undergoing the charge-exchange collision reflecting the details of the potential mixing of the incoming and outgoing channels. We measure $A$ for the $^{40}$Ca$^+$ ions in the ground $S_{1/2}$ state and the excited $P_{1/2}$, $D_{3/2}$ and $D_{5/2}$ states. 
We also performed the calculation of the LiCa$^+$ potential-energy curves, and from the comparison between the experimental results and theory calculations, we have identified the route of the charge-exchange collisions.
A profound understanding of the interaction potentials is an important step to predict the route of molecular-ion formation, 
which may allow us to realize the creation of translationally cold molecules composed of a large number of atoms and to reveal the physics of the growth of mesoscopic molecular ions \cite{Cote2}.


\section{\label{sec:level2}Experimental setup}
\begin{figure}[tb]
\includegraphics[clip,width = 8.5cm]{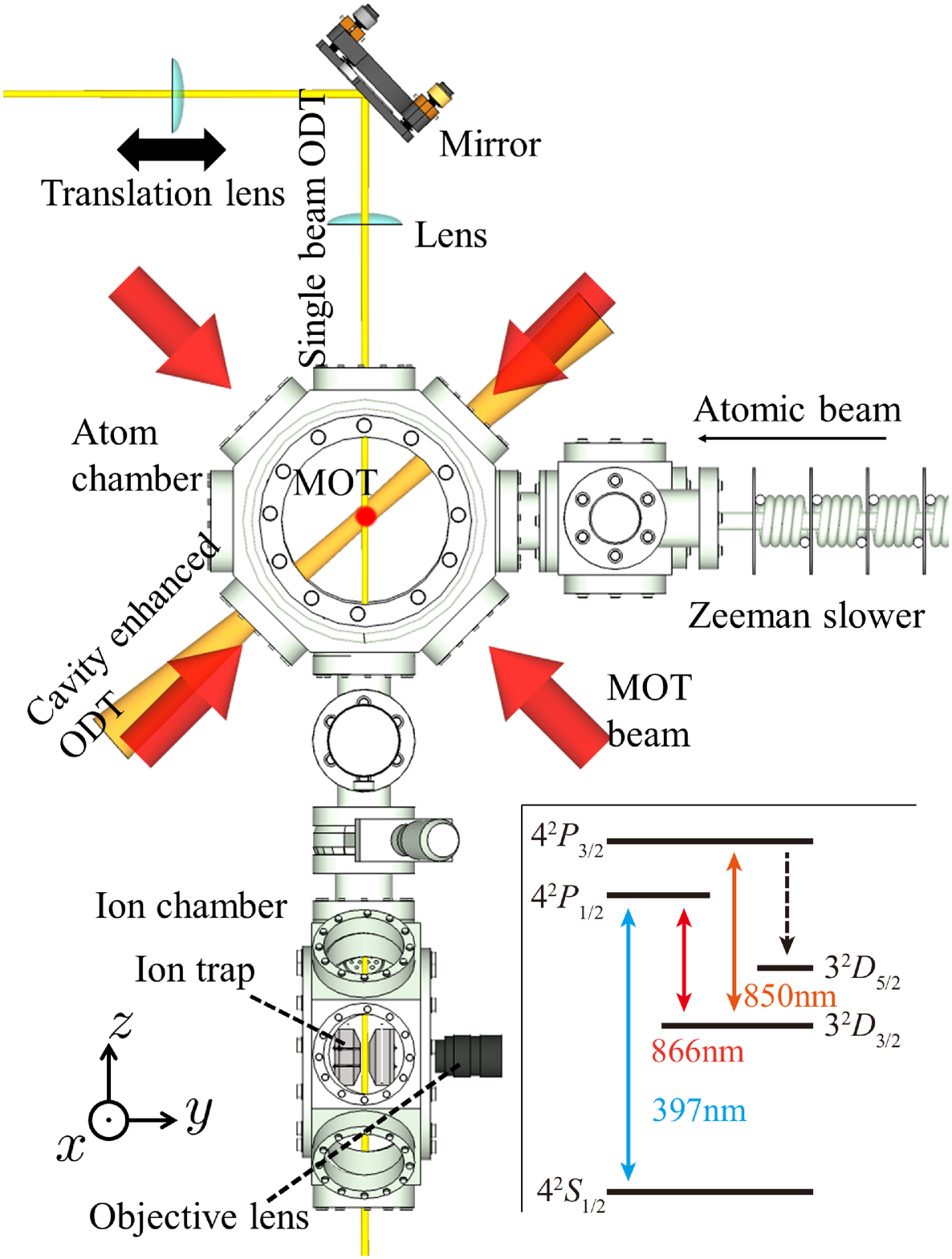}
\caption{\label{fig:system} Schematic drawing of the experimental setup. The upper part shows the atom chamber whereas the lower part presents the ion chamber. 
$\rm{^6Li}$ atoms are supplied from the right side of the atom chamber and are decelerated by a Zeeman slower. First, $\rm{^6Li}$ atoms are trapped by a conventional MOT, and, then, they are trapped by the cavity-enhanced trap and transferred into an ODT at the center of the atom chamber. Ca$^+$ ions are trapped at the center of the ion-trap electrode placed inside the ion chamber. 
The 397-nm cooling laser for the ions is incident along both the radial and the axial directions of the ion trap, and the 866-nm repump laser is incident along the axial direction. Fluorescence of the ions is collected by an objective lens and detected by a PMT and EMCCD camera. The enegy-level diagram of the $^{40}$Ca$^+$ ion is shown in the inset.}
\end{figure}

In our experiment, $^6$Li atoms and $\rm{^{40}Ca^+}$ ions were trapped and cooled separately in a different part of a vacuum chamber as shown in Fig. \ref{fig:system}. After the preparation of the atoms and the ions, the atoms were transferred to the center between the ion trap electrodes using an optical tweezer, and mixed with the ions. The ions were trapped by a conventional linear Paul trap with an rf voltage with an amplitude of 44~V and frequency $\Omega = 2\pi \times 4.8$~$\rm{MHz}$, realizing the radial and the axial trap frequencies of $(\omega_r, \omega_z) = 2 \pi \times (252, 95)$~kHz, respectively. 
Neutral $\rm{^{40}Ca}$ atoms from the oven were photoionized using 423~nm and 375~nm lasers. Then the ions were cooled with a 397~nm laser, whose frequency was red-detuned from the $S_{1/2}-P_{1/2}$ transition and with a 866~nm laser, as a repump light, tune to the resonance of the $D_{3/2}-P_{1/2}$ transition. Fluorescence from the ions was detected by a photomultiplier tube (PMT) and an electron multiplying CCD (EMCCD) camera. The ion trap system included compensation electrodes to minimize the excess micromotion of the trapped ions. Minimization of the micromotion was accomplished by the RF photon correlation method, and the residual micromotion energy at the optimum condition was estimated to be $k_{\rm B} \times$ 0.2 ${\rm mK}$. 

${\rm^6Li}$ atoms were first trapped by a conventional magneto-optical trap (MOT) and then transferred into a cavity-enhanced optical dipole trap (ODT) \cite{Mosk}. Eventually, they were transferred into a single optical dipole trap created by a 1064-nm laser \cite{Inada}. 
Then, the atoms were transported to the position of the ion trap to create the atom-ion mixture by moving one lens placed on the translation stage. The trap frequencies for the atoms were 670~Hz in the radial direction and 1.7~Hz in the axial direction. The peak density of the atoms was $(4.8 \pm 0.3) \times 10^8$~$\rm{cm^{-3}}$. The temperature of the ${\rm^6Li}$ atoms was $(5.8 \pm 0.1)$ $\rm{\mu K}$. ${\rm^6Li}$ was prepared in the ground hyperfine state $F=1/2$ of $^2S_{1/2}$ when it was mixed with the $\rm{^{40}Ca^+}$ ions in the current experiment.

\section{\label{sec:level3-1}Calibration of the collision energy}
Under the condition that the kinetic energy of the ion is much larger than that of the atoms, the atom-ion collision energy  $E_{\rm{coll}}$ is described by the kinetic energy of the ion $E_{\rm{ion}}$ as
\begin{eqnarray}
E_{\rm{coll}} = \frac{\mu}{m_{\rm{ion}}}E_{\rm{ion}}\label{eq:collision_energy},
\end{eqnarray}
where $\mu$ is the reduced mass of the $\rm{^6Li}-\rm{^{40}Ca^+}$ combination. Since $\mu/m_{\rm{ion}} = 0.13$ in our case, the collision energy $E_{\rm{coll}}$ is considerably smaller but proportional to $E_{\rm{ion}}$. In this work, we tune $E_{\rm{coll}}$ through the tuning of $E_{\rm{ion}}$ due to deliberate excitation of the excess micromotion in a controlled way explained below.

\begin{figure}[tb]
\includegraphics[clip,width = 7.5cm]{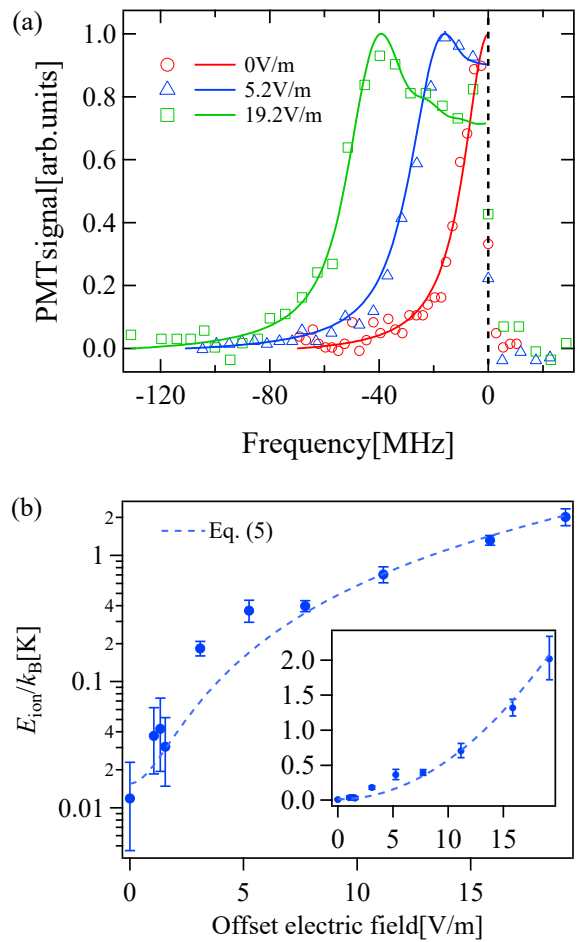}
\caption{\label{fig:energy}
(a) Normalized fluorescence spectra of the Ca$^+$ ions for different offset electric fields. Horizontal and vertical axes show the cooling laser detuning and the fluorescence intensity of the ions, respectively. (b) The total kinetic energy of the ion determined from the spectra vs. the offset electric field. Error bars indicate the standard errors. The dashed curve shows the relation described by Eq. (\ref{eq:quadratic}) taking $E_{\rm{min}}$ as a fitting parameter. The inset shows the same plot with the vertical axis shown in a linear scale.}
\end{figure}

We start with the condition in which the micromotion of the ions is well compensated, and then we apply an offset electric field to displace the ions in the radial direction ($x$ and $y$ directions) to intentionally add excess micromotion energy to the ions. The added kinetic energy of the ions is determined from the shape of the ions' fluorescence spectra.
When the micromotion is excited, the spectral shape gets broadened and deformed, as shown in Fig. \ref{fig:energy}(a). The spectral intensity $P$ can be described as
\begin{eqnarray}
P=\frac{\chi^2}{4}\sum_{n=-\infty}^{\infty}{\frac{J_n^2(\beta)}{\left(\Delta\omega+n\Omega\right)^2+\left(\gamma/2\right)^2,}}\label{eq:spectrum}
\end{eqnarray}
where $\Delta\omega$, $\gamma$, and $\chi$ are the laser detuning, the natural linewidth of the $S_{1/2}-P_{1/2}$ transition, and the electric dipole moment of the transition, respectively, and $J_n$ is the $n$-th order Bessel function for $\beta$ \cite{Devoe, Blumel, Berkeland}. 
The value $\beta$ is the modulation index defined by $\beta = |\frac{kv_{\rm{r}}\cos\theta}{\Omega}|$, where $k$, is the wavenumber of the cooling laser, $v_{\rm{r}}$ is the velocity of the ion along the direction of the micromotion (namely, the  $x$ and $y$ directions in our case), and $\theta$ is the angle of the propagation direction of the cooling laser to the $x$-$y$ plane. 
When $\beta=0$, the Bessel function can only be $n = 0$, and the spectral shape becomes a simple Lorenz function. When $\beta$ has a finite value, the spectral shape is deformed and shows a peak around $| \Delta\omega | \approx \beta\Omega$. We determine $\beta$ by fitting the spectra with Eq. (\ref{eq:spectrum}). 

Figure~\ref{fig:energy}(a) shows the fluorescence intensity of the Ca$^+$ ions measured by the PMT plotted against cooling laser detuning. The red, blue and green symbols show the fluorescence spectra obtained at an applied electric field strength of 0, 5.2, and 19.2~V/m, respectively.
The background PMT signal intensities due to stray light are subtracted from the spectra. After the background subtraction, the fluorescence spectra are normalized to be from zero to unity.
We fit the spectra with Eq. (\ref{eq:spectrum}) taking $\beta$ as a free parameter. The resonance position is chosen to be at the frequency where the fluorescence from the ion drastically decreases. The solid red, green and blue curves show the results of the fitting with Eq. (\ref{eq:spectrum}), with $\beta=0.73, 5.0, 10$ for the red, blue and green data, respectively. 

The excess micromotion energy of the ion $E_{\rm{emm}}$ is determined by $\beta$ from  the expression \cite{Berkeland}, 
\begin{eqnarray}
E_{\rm{emm}} = \frac{m_{\rm{ion}}\Omega^2}{2k^2{\cos^2{\theta}}}\beta^2,\label{eq:emmbeta}
\end{eqnarray}
where $m_{\rm{ion}}$ is the mass of the Ca$^+$ ion.
The total kinetic energy of the Ca$^+$ ion is 
\begin{eqnarray}
E_{\rm{ion}}=E_{\rm{min}}+E_{\rm{emm}},
\label{eq:total_energy}
\end{eqnarray}
where the first term on the right-hand side is the minimum ion energy we can reach in the experiment.

Figure \ref{fig:energy}(b) illustrates the kinetic energy of the ion determined from the fluorescence spectra plotted against the applied offset electric field. 
The electric field is enforced by applying the voltage to the compensation electrode, and the electric field strength is calibrated through the displacement of the ions measured using the CCD camera. The dashed curve shows the  quadratic dependence of $E_{\rm{emm}}$ on the applied electric field which can be described as \cite{Harter}
\begin{eqnarray}
E_{\rm{emm}} = \frac{4}{m_{\rm{ion}}} \left( \frac{e\mathcal{E}_r}{q\Omega} \right)^2\label{eq:quadratic},
\end{eqnarray}
where $e$, $\mathcal{E}_r$ and $q$ are an elementary charge, the applied electric field strength, and the stability parameter of the ion trap determined from the secular frequency measurements ($q = 0.15$), respectively \cite{Berkeland, Harter, Meir}.
The dashed curve shows the relation described by Eq. (\ref{eq:quadratic}) taking $E_{\rm{min}}$ as a fitting parameter. $E_{{\rm min}}/k_{\rm B} \sim$10~mK is limited by the imperfection of the laser cooling condition.
The experimental data points and the expected quadratic curve show great consistency, and therefore we have been able to experimentally confirm the validity of Eq. (\ref{eq:quadratic}) for the ion energy calibration.
The dashed curve is used to interpolate the data for the determination of the energy of the ion at the intermediate electric-field strength.


\section{\label{sec:level3-2}Energy dependence of the charge-exchange collision cross sections}

The inelastic collisions that we observe in our system have been identified as charge-exchange collisions through the mass spectrometry of the reaction-product ions in our previous work \cite{Haze2}. In the current experiment, three to seven Ca$^+$ ions are loaded in the ion trap to form a one-dimensional crystal. Then ions are mixed with Li atoms and held for one second, and we count the number of ions lost from the trap. 
Within 1~s of the holding, the number of ions decays in time. We repeat the measurement 50 times, and the ion-loss probability $P_{\rm CE}$ as a function of holding time is measured.
We derive the charge-exchange collision rate $\Gamma_{\rm CE}$ from the solution of the rate equation which takes into account the radiative decay of the ion from the metastable $D_{3/2}$ and $D_{5/2}$ states, $P_{\rm CE} = \frac{\Gamma}{\gamma+\Gamma} \mathrm{e}^{-(\gamma+\Gamma)t} + \frac{\gamma}{\gamma+\Gamma}$, where $\gamma$ is the spontaneous decay rate of the $D_{3/2}$ and $D_{5/2}$ states and $t$ is the mixing time of the atoms and ions \cite{Haze2}.
The energy-dependent charge-exchange collision cross section $\sigma_{\rm CE}(E)$ can be derived from the relation $\sigma_{\rm CE}(E) = \Gamma_{\rm CE} / n v$, where $n$ is the atomic density and $v$ is the mean relative velocity of the colliding atom and ion.
The atomic density $n$ at the ion position is determined from the information on the number of atoms, the trap frequencies for the atoms, and the displacement of the ion position from the center of the atomic cloud. 
Although an elastic collision between a Li atom and a Ca$^+$ ion causes the loss of the atoms, the reducation of the number of atoms in the current experiment is negligible. The relative velocity of the Li atom and the Ca$^+$ ion is determined solely by the velocity of the ion, since the velocity of the ion is much greater than the velocity of the atom. 

\begin{figure}[b]
\includegraphics[clip, width= 8.2cm]{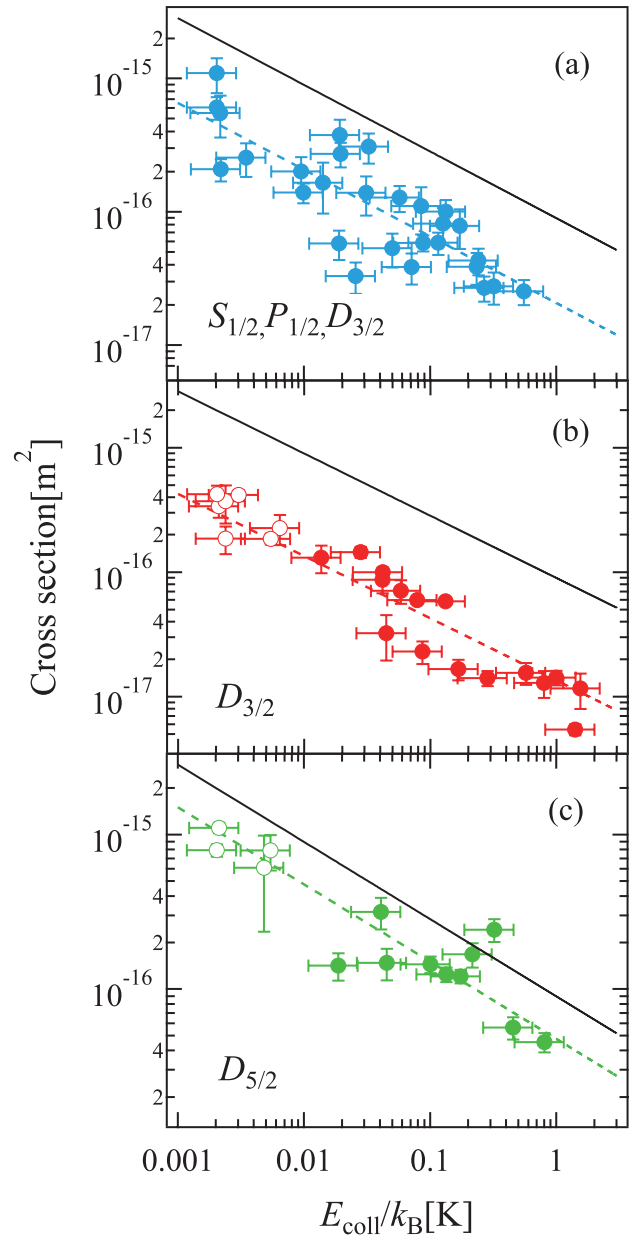}
\caption{\label{fig:cross} Charge-exchange collision cross sections as a function of the collision energy for the $\rm{Ca^+}$ ions in (a) the mixture of the $S_{1/2}$, $P_{1/2}$ and $D_{3/2}$ states, (b) the $D_{3/2}$ state, and (c) the $D_{5/2}$ state, respectively.
The data shown with the open circles are the points which may suffer from the ion heating due to rf noise, and are also excluded from further analysis. The black solid lines indicate the Langevin cross section given by Eq. (\ref{eq:Langevin}). The dashed lines show the fitting result of the data with the $E^{-1/2}$ dependence predicted from the Langevin model.}
\end{figure}

Figure \ref{fig:cross}(a) shows the measured charge-exchange collision cross sections as a function of the collision energy $E_{\rm{coll}}$ for the $\rm{Ca^+}$ ions under the irradiation of the cooling and repump lasers. In this condition, ions are considered to be in the mixture state of the $S_{1/2}$, $D_{3/2}$ and $P_{1/2}$ states. The solid line in Fig. \ref{fig:cross}(a) is the energy-dependent Langevin collision cross section expressed as \cite{langevin1905formule}
\begin{eqnarray}
\sigma_{\rm L} (E_{\rm{coll}}) = \pi \sqrt{\frac{2C_4}{E_{\rm{coll}}}} \approx (3.3 \times 10^{-28}) E_{\rm{coll}}^{-1/2}  [{\rm m^2}]
\label{eq:Langevin}
\end{eqnarray}
for the $^6$Li--$^{40}$Ca$^+$ combination, where $C_4=\alpha Q^2/4\pi\epsilon_0$, with $\alpha$, $Q$ and $\epsilon_0$ being the static polarizability of the $^6$Li atoms, the charge of the $\rm{Ca^+}$ ion and the vacuum permittivity, respectively. 
Since the charge-exchange collision happens when the atom-ion distance is very small, the charge-exchange collision is part of the Langevin collision and therefore $\sigma_{\rm CE} (E_{\rm{coll}})$ is proportional to but smaller than $\sigma_{\rm L} (E_{\rm{coll}})$. The energy dependence of the data shown in Fig. \ref{fig:cross}(a) is consistent with that of the Langevin model in the collision energy range of our measurements.

In order to reveal the internal-state dependence of the charge-exchange collision cross sections,
we prepare $\rm{^{40}Ca^+}$ ions in the various internal states using an optical pumping.  
To prepare the ion in the ground $S_{1/2}$ state, we turn off the cooling laser 20ms earlier than the repump laser.
To prepare the ion in the $D_{3/2}$ state, the repump laser is turned off 20ms earlier than the cooling laser.
To prepare the ion in the $D_{5/2}$ state, we turn off the repump laser, and then the 850-nm laser that drives the ions from $D_{3/2}$ to $P_{3/2}$ is irradiated together with the cooling laser for 20 ms.
The pumping of the ions is done just before the atoms are transported to the ion-trap position. We have confirmed that the fluctuation of the transportation timing does not affect our measurement of the charge-exchange rate.

Figures \ref{fig:cross}(b) and \ref{fig:cross}(c) show the charge-exchange collision cross sections vs $E_{\rm{coll}}$ for the $\rm{Ca^+}$ ions prepared in the metastable $D_{3/2}$ and $D_{5/2}$ states.
In addition, in contrast to the case of ions under cooling lasers, ions prepared in the metastable $D$ states are not irradiated by the cooling lasers, and therefore ions may be exposed to heating by an rf noise \cite{RevModPhys.87.1419, Wineland98experimentalissues} and the coupling between the micromotion and the secular motion in the presence of multiple ions \cite{Berkeland}.
The rate of the heating due to the rf noise depends on the spectral density of the electric-field noise, which obeys the empirically obtained power law to the ion-electrode distance \cite{RevModPhys.87.1419, Wineland98experimentalissues}. We estimate the rate of the heating due to rf noise using the typical value of the spectral density from the power law, and it is estimated to be 70 mK/s for the geometric layout of our ion trap. Since our measurement of the charge-exchange cross section may suffer from the heating due to the rf noise, we exclude the data of the collision energy less than 10 mK [ shown with the open symbols in Figs. \ref{fig:cross}(b) and(c)] from the analysis of energy dependence of the cross sections provided below.
On the other hand, a precise determination of the heating rate due to the micromotion-secular-motion coupling is rather an open issue. However, this heating effect does not seriously affect our conclusion \cite{footnote}. Therefore this coupling effect is not taken into account in our analysis.

We fit the data shown in Figs. \ref{fig:cross}(a)-\ref{fig:cross}(c) to the function $\sigma_{\rm CE} (E_{\rm{coll}})=A\sigma_{\rm L} (E_{\rm{coll}})$ with the pre-factor $A$ as a free parameter. 
The dashed lines in Figs. \ref{fig:cross}(a)-\ref{fig:cross}(c) are the fitting result to the data. The obtained values of $A$ are summarized in Table \ref{tab:table1}. To derive $A$ for the $P_{1/2}$ state, we estimate the population distribution of the ion in the $S_{1/2}$, $P_{1/2}$, and $D_{3/2}$ states. From the information on the intensity and the detuning of the cooling and the repump lasers, the population distribution is determined to be $(p_S, p_P, p_D) = (0.42, 0.29, 0.29)$. Since the collision cross section for the mixed state is described as $\sigma_{\rm{mixed}} = p_S \sigma_S + p_P \sigma_P + p_D \sigma_D$, the $A$ coefficient for the $P_{1/2}$ state is determined from $A$ for the $S_{1/2}$ and $D_{3/2}$ states. As for the ions in the $S_{1/2}$ state, the charge-exchange collision rate is too small to be measured accurately in the current experiment. We have determined the upper limit of $A$ for the $S_{1/2}$ state to be $1 \times 10^{-3}$, which corresponds to our measurement limit of the smallest charge-exchange rate of $10^{-3}$ s$^{-1}$.\\

\begin{table}[t]
 \caption{\label{tab:table1}
$A$ for various internal states of $^{40}$Ca$^+$ ions. }
\begin{ruledtabular}
\begin{tabular}{ll}
$\rm{^{40}Ca^+}$ internal state&$A$\\
\colrule
$S_{1/2}$&$\alt1\times10^{-3}$\\
$D_{3/2}$&$0.15\pm0.01$\\
$D_{5/2}$&$0.53\pm0.06$\\
Mixed state of $S_{1/2}$, $P_{1/2}$, and $D_{3/2} $&$0.23\pm0.03$\\
$P_{1/2}$&$0.64\pm0.04$\\
\end{tabular}
\end{ruledtabular}
\end{table}


\section{\label{sec:theory}Calculation of the potential energy curves}

\begin{figure}[tb]
\includegraphics[clip, width= 8.7cm]{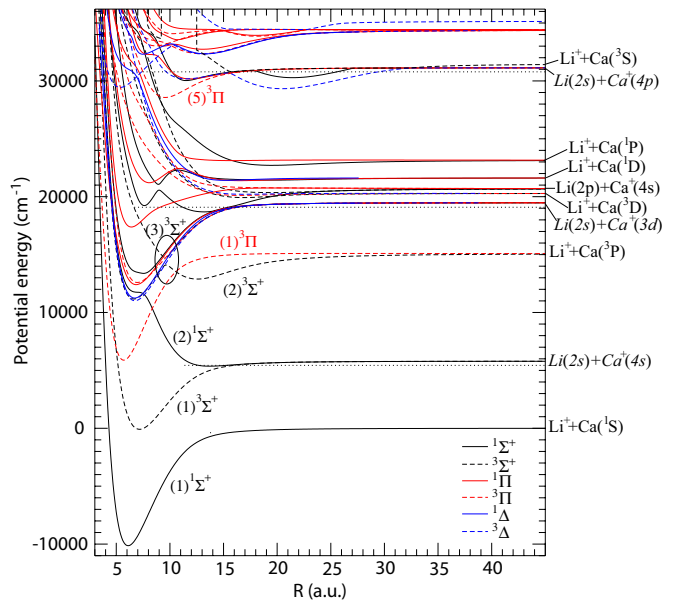}
\caption{ LiCa$^+$ potential energy curves computed in the present work relevant for the experiment. The circle locates the avoided crossing which is expected to be active for the charge-exchange process starting from Li($2s$)+Ca$^+$($3d$) for the $^3\Sigma^+$ symmetry. \label{fig:PEC}}
\end{figure}

The knowledge of the molecular properties of the LiCa$^+$ system yields useful insight to interpret the results above. Surprisingly enough, no information was available in the literature for this system up to now. Following the approach developed in Refs. \cite{aymar2011,aymar2012}, we calculated the LiCa$^+$ potential-energy curves (PECs) for all symmetries up to the Li($2s$)+Ca$^+$($4p$) dissociation limit, namely, the ninth asymptote. Computing such highly excited states is indeed easily tractable with our approach. Generally speaking, a diatomic molecules composed of an alkali-metal atom and an alkaline-earth-metal ion can be modeled as a system where the two valence electrons are moving in the field of the two ionic cores Li$^+$ and Ca$^{2+}$. The latter are represented by two effective core potentials which include core polarization effects. The Schr\"odinger equation for the two valence electrons can then be solved in a configuration space spanned by a large Gaussian basis set. 
With only two active electrons, a full configuration interaction can thus be achieved \cite{aymar2011,aymar2012}, yielding the ground-state PEC and numerous excited-state PECs, displayed in Fig. \ref{fig:PEC}.

In contrast to heavier systems like RbCa$^+$ and RbBa$^+$ \cite{Hall,Hall2,Hall3}, the LiCa$^+$ species is light enough to allow accessing to the PECs up to Li($2s$)~+~Ca$^+$($4p$), thus representing an appealing species to sort out the complicated dynamics taking place in such a hybrid trap. 
Assuming that only two-body reactions are expected in the present experimental conditions, one can immediately infer several statements allowing us to understand the present experimental results. First there is a strong avoided crossing (circled in Fig. \ref{fig:PEC}) between the (2)$^3\Sigma^+$ and (3)$^3\Sigma^+$ PECs which is probably responsible for the large observed charge-exchange rate in the Li($2s$)~+~Ca$^+$($3d$) entrance channel. As it is well isolated from other curves of the same symmetry, this avoided crossing can be linearized following the standard Landau-Zener approach, so that a single path transition probability can be estimated according to $P_{\rm LZ}=\exp(-2 \pi W_c^2/(v_c \Delta F_c))$. In this expression the interaction strength parameter $W_c=0.0015025$~a.u. is half of the potential energy spacing between the two PECs at the crossing point located at $R_c=9.69a_0$ ($a_0$ is the Bohr radius).
The classical relative velocity $v_c=\sqrt{2U_c/\mu}=0.002082$~a.u. at $R_c$ of the particles with reduced mass $\mu$ is obtained through the value $U_c=0.023533$~a.u. of the potential energy at $R_c$ with respect to the  Li($2s$)~+~Ca$^+$($3d$), assuming a negligible incoming velocity at large distance. Finally, the $\Delta F_c=0.011612$~a.u. is the difference in slopes between the two branches of the linearized crossing. Altogether, we obtain $P_{\rm LZ}=0.556$, which is large enough to explain the large observed charge-exchange rate. Note, however, that several further issues should be taken in account for a more quantitative estimate. First, assuming a statistical population of the initial molecular states, the (3)$^3\Sigma^+$ state would be populated with a weight of 3/20.
On the other hand, spin-orbit interaction probably plays a significant role in the dynamics; in particular, the (2)$^3\Pi$, (1)$^1\Pi$, and (1)$^3\Delta$ state are expected to be efficiently coupled to the (3)$^3\Sigma^+$ state, so that the effective incoming flux toward the avoided crossing is probably higher than the one inferred by the statistical weight. Moreover, the inclusion of the spin-orbit interaction in the model would allow us to interpret the observed difference of rates for the $D_{3/2}$ and $D_{5/2}$ states.

The reported almost unmeasurable rate for the charge-exchange reaction starting from the Li($2s$)~+~Ca$^+$($4s$) entrance channel is also readily interpreted from the behavior of the PECs. Just like in other similar species \cite{Humberto}, two molecular states are correlated to this asymptote, namely, the (1)$^3\Sigma^+$ and the (2)$^1\Sigma^+$. The former state is the lowest one in this symmetry, so that only elastic collisions are expected in this channel. The latter state lies far above the LiCa$^+$ electronic ground state, so that no clear efficient avoided crossing is present. Therefore the only possible two-body reaction proceeds through spontaneous emission, leading to either radiative charge exchange leading to two ground state species (a Li$^+$ ion and a neutral Ca atom) or to the formation of a ground-state molecular ion LiCa$^+$, as already reported in Ref. \cite{Hall}.
Following the method exposed in \cite{Humberto}, we estimated that the cross section for these radiative processes is about 50 times smaller than the one calculated for RbCa$^+$, namely, smaller than 10$^{-17}$~cm$^2$. It is likely that this process could be observed provided that an improved detection using mass spectrometry is implemented. 

Finally, the situation is more involved for the Li($2s$)~+~Ca$^+$($4p$) entrance channel. This asymptote is quite well isolated from neighboring ones, so that there is no obvious avoided crossing in the relevant PECs. However, we see that at almost vanishing initial energy (see the horizontal dotted line in Fig.\ref{fig:PEC}) the (5)$^3\Pi$ PEC crosses several curves correlated to lower asymptotes to which it can be coupled through spin-orbit interaction, thus allowing for efficient charge exchange. An estimate of the transition probability would require a model for the molecular spin orbit for these states, and thus further quantum chemistry calculations, which are out of the scope of the present paper.


\section{\label{sec:level4}Conclusion}
In conclusion, we have systematically studied the energy dependence and the internal-state dependence of the charge-exchange collision cross sections for the $^6$Li -- $^{40}$Ca$^+$ combination. The energy dependence is confirmed to be consistent with the Langevin model in the collision energy range realized in this work.
The inelastic collision rate depends on the internal state in which the ions are prepared, and the direct comparison between experiment and theory enables us to pin down the route of the charge-exchange process and also to confirm the calculation result of the potential curves with the experiment. Considering that the $^6$Li--$^{40}$Ca$^+$ combination has a quite low heating rate arising from a micromotion \cite{Marko}, the $^6$Li--$^{40}$Ca$^+$ combination is one of the most promising candidates to investigate the quantum collisions in an atom-ion hybrid system. The detailed understanding of the inelastic collisions and interaction potentials will be an important step toward the realization of the full quantum control of the elementary processes of chemical reactions in an ultracold-temperature regime.

\section{\label{sec:level4}ACKNOWLEDGMENTS}
This work is partially supported by JSPS KAKENHI  (Grants No. 26287090, No. 24105006, No. 15J10722, and No, JP16J00890), and PRESTO, JST.
H.D.S Jr, M.R., and O.D. acknowledge the support of the Marie-Curie Initial Training Network ‘COMIQ: Cold Molecular Ions at the
Quantum limit’ of the European Commission under the Grant Agreement No. 607491. R. Vexiau is gratefully acknowledged for double-checking the LiCa$^+$ PECs.

{\it Note added.} — Recently, we became aware of
a closely related work on the calculation of the LiCa$^+$ PECs by Habli {\it et al} \cite{Habli}. Their method is very close to ours, as attested to by several common references. No detailed comparison was performed as no supplementary data were provided with Ref. \cite{Habli}, but a look at the figure for PEC curves reveals an overall good agreement between our calculations.


\begin{thebibliography}{90}
\bibitem{footnote} If we assume a factor of five larger heating for the data in Fig. \ref{fig:cross}(b) and \ref{fig:cross}(c), we still have data points enough to determine the energy dependence of the charge-exchange collisions, and we obtain the same results in the pre-factor $A$. Therefore, we conclude that the extra heating due to the micromotion-secular-motion coupling does not change the conclusion of our results.
\end{thebibliography}
\end{document}